\documentclass[fleqn,10pt,a4paper]{article}
\usepackage{amsmath,amssymb,bm}
\usepackage{cite,epsfig,color,graphicx}

\title{Longitudinal wave-breaking limits in a unified geometric model of relativistic warm plasmas}
\author{DA Burton\\
Department of Physics,\\
Lancaster University, UK\\
and The Cockcroft Institute, UK
\and
A Noble\\
Department of Physics,\\
Lancaster University, UK\\
and University of Strathclyde,
UK\\
}
\begin{document}
\maketitle
\begin{abstract}
The covariant Vlasov-Maxwell system is used to study breaking of
relativistic warm plasma waves. The well-known theory of relativistic
warm plasmas due to Katsouleas and Mori (KM) is subsumed within a unified
geometric formulation of the `waterbag' paradigm over
spacetime. We calculate the maximum amplitude $E_\text{max}$ of
non-linear longitudinal electric waves for a particular class of
waterbags whose geometry is a simple $3$-dimensional generalization (in velocity) of the
$1$-dimensional KM waterbag (in velocity). It is well known that the value of $\lim_{v\rightarrow
  c}E_\text{max}$ (with the effective temperature of
the plasma electrons held fixed) diverges for the KM
model; however, we show that a certain class of simple $3$-dimensional
waterbags yields a finite value for $\lim_{v\rightarrow c}E_\text{max}$, where $v$ is the phase velocity
of the wave and $c$ is the speed of light.
\end{abstract}
\section*{Introduction}
Considerable effort has been devoted to developing compact
accelerators employing the enormous electric fields present in
plasma wakes driven by intense lasers~\cite{tajima:1979} or charged
particle beams~\cite{chen:1985} (see~\cite{malka:2008, caldwell:2009}
for recent discussions). Conventional accelerators operate
by exciting RF microwave cavities with klystrons and use the
longitudinal electric component of a cavity mode to accelerate bunches of charged
particles for subsequent collision. However, it is anticipated that
electric field strengths in the
next generation of accelerators will be so high that the RF cavity
walls may undergo electrical breakdown~\cite{wuensch:2002}. To address this issue,
researchers have turned to plasma-based acceleration mechanisms whose 
field can be orders of magnitude beyond that of
conventional accelerators. Recent years have seen the on-going development of
\emph{compact} sources of intense electromagnetic radiation in the
X-ray to THz frequency range~\cite{schlenvoigt:2008} that employ
laser-driven plasma acceleration. Such sources promise a wide range
of applications in medicine, material science and security.   

A sufficiently short and intense laser pulse propagating
through a plasma may create a travelling longitudinal plasma
wave whose velocity is approximately the same as the laser pulse's group
velocity. However, it is not possible to sustain arbitrarily large
electric fields; substantial
numbers of plasma electrons become trapped in the wave and are
accelerated, which dampens the wave. Indeed, the trapping
phenomenon in longitudinal plasma waves lies at the heart of the
original laser wakefield accelerator concept~\cite{tajima:1979}.   

Although the evolution of a plasma wave dynamically
trapping particles is complex, over the years much
effort has been devoted to analytically understanding the upper bound
(`wave-breaking limit') on the amplitude of plasma waves.
Wave-breaking limits were first calculated for cold
plasmas~\cite{dawson:1959, akhiezer:1956} undergoing non-linear
longitudinal electrostatic
oscillations, and thermal effects were later included in
non-relativistic~\cite{coffey:1971} and relativistic~\cite{katsouleas:1988, rosenzweig:1988, schroeder:2005}
contexts. The results for the cold plasma are uncontroversial, but
recent discussion~\cite{trines:2006, schroeder:2007, trines:2007} has
uncovered
difficulties with establishing an agreed analytical description of longitudinal
wave-breaking in warm plasmas; in particular, it has been noted that
different plasma models based on different assumptions yield different
results. Models of non-linear plasma
waves near breaking are approaching the limits of their
domain of applicability, and different models exhibit different wave-breaking limits.
Although
recent experiments~\cite{mangles:2004, geddes:2004, faure:2004}
operate in the 3-dimensional `bubble' (or `blow-out')
regime~\cite{lu:2006} and exploit \emph{transverse}
wave-breaking~\cite{esirkepov:2006}, recent work~\cite{trines:2006,
  schroeder:2007, trines:2007} has rekindled interest in the theory of
longitudinal wave-breaking.

Recent discussion~\cite{trines:2006, schroeder:2007, trines:2007} includes
comparison of the behaviour of the relativistic `waterbag'
model~\cite{katsouleas:1988, mori:1990} due to Katsouleas and Mori (abbreviated as KM) and a
warm plasma model~\cite{schroeder:2005} due to Schroeder,
Esarey and Shadwick (abbreviated as SES) employing velocity moments of
the $1$-particle plasma electron distribution.
The KM and SES models yield different results for the maximum
amplitude of non-linear electrostatic oscillations 
in the limit $v\rightarrow c$ with
the temperature of the plasma held fixed ($v$ is the phase velocity of the plasma
wave with respect to the laboratory frame). The KM maximum electric
field diverges logarithmically in $\gamma=1/\sqrt{1-v^2/c^2}$ as $v\rightarrow
c$, whereas the SES maximum electric field tends to a finite value as
$v\rightarrow c$ (with the initial plasma temperature held fixed in
the limit $v\rightarrow c$). Employing velocity moments of the 1-particle plasma electron
distribution, SES require that the distribution remains narrow
in velocity spread whereas the KM approach employs a particular waterbag
solution to the Vlasov equation. The disagreement of the two
approaches has been attributed to the waterbag's piecewise constant
structure and lack of a tail~\cite{schroeder:2007}.

The KM model is formulated over $2$-dimensional spacetime and SES employ a line
distribution in longitudinal velocity to simplify their field
equations on $4$-dimensional spacetime. Neither model admits a plasma
electron distribution with a non-vanishing transverse thermal
spread. Thus, a theory of waterbags over $4$-dimensional spacetime was
recently developed~\cite{burton_noble:2009, burton_noble_wen:2009} to permit
analytical investigation of wave-breaking as a function of the
$3$-dimensional shape (in velocity) of the plasma electron
distribution.
In the following we cast the KM field equations in a form comparable with those
of the waterbag over $4$-dimensional spacetime and, for the first time, give a unified
presentation of the derivation of wave-breaking limits in the KM model
and our waterbag model. We conclude with a comparison of
the predictions of a particular class of our
waterbags, the KM model and SES model. We find that
the results of our present approach have more in common with the SES
model than the KM model.
\section{Vlasov-Maxwell equations}
The brief summary of the Vlasov-Maxwell
equations given below establishes our conventions; further details
may be found in~\cite{burton_noble_wen:2009, ehlers:1971}. We employ
the Einstein summation convention throughout this article.
Latin indices $a,b,c$ run over $0,1,2,3$ and units are used in which the speed of light
$c=1$ and the permittivity of the vacuum $\varepsilon_0=1$.

Let $(x^a)$ be an inertial coordinate system on Minkowski spacetime
$(\mathcal{M},g)$ where $x^0$ is the proper
time of observers at fixed Cartesian coordinates $(x^1,x^2,x^3)$ in
the laboratory. The metric tensor $g$ has the form
\begin{equation}
\label{metric}
g = \eta_{ab}\, dx^a \otimes dx^b
\end{equation}
where 
\begin{equation}
\eta_{ab} =
\begin{cases}
&-1\text{   if $a=b=0$},\\
&1\text{    if $a=b\neq 0$},\\
&0\text{    if $a \neq b$}.
\end{cases}
\end{equation}

Let $(x^a,\dot{x}^b)$ be an induced coordinate system
on the total space $T\mathcal{M}$ of the tangent bundle
$(T\mathcal{M},\Pi,\mathcal{M})$ and
in the following, where convenient, we will write $x$ instead of $x^a$ and $\dot{x}$
instead of $\dot{x}^b$. For notational simplicity, we will not
distinguish between a point in a manifold and its coordinate representation.
 
The total space $\mathcal{E}$ of the sub-bundle
$(\mathcal{E},\Pi|_{\cal E},\mathcal{M})$ of $(T\mathcal{M},\Pi,\mathcal{M})$ is the
set of timelike, future-directed, unit normalized tangent vectors on $\mathcal{M}$,
\begin{equation}
\mathcal{E}=\{(x,\dot{x})\in
T\mathcal{M}\, \big|\, \varphi = 0\,\,\text{and}\,\, \dot{x}^0>0\}
\end{equation}
where
\begin{equation}
\varphi = \eta_{ab}\,\dot{x}^a\dot{x}^b + 1.
\end{equation}

Plasma electrons are described statistically by a $1$-particle
distribution $f$ on $T{\cal M}$ which induces a number $4$-current vector
field $N$,
\begin{align}
&N = N^a \frac{\partial}{\partial x^a},\\
\label{component_number_current_forms}
&N^a(x) = \int_{\mathcal{E}_x} \dot{x}^a f\,\iota_X\# 1,
\end{align}
where $\mathcal{E}_x = (\Pi|_{\cal E})^{-1}(x)$ is the fibre of
$(\mathcal{E},\Pi|_{\cal E},\mathcal{M})$ over $x\in\mathcal{M}$. The
$3$-form $\iota_X\# 1$ on $T{\cal M}$ is induced from the $4$-form $\# 1$,
\begin{equation}
\label{definition_hash_1}
\# 1 = d\dot{x}^0 \wedge d\dot{x}^1 \wedge d\dot{x}^2 \wedge d\dot{x}^3,
\end{equation}
and the dilation vector field $X$,
\begin{equation}
\label{definition_X}
X = \dot{x}^a\frac{\partial}{\partial{\dot{x}^a}},
\end{equation}
on $T\mathcal{M}$, where $\iota_X$ is the interior product on forms. It may be shown
\begin{equation}
\iota_X \# 1 \simeq \frac{1}{\sqrt{1+|\dot{\bm{x}}|^2}}\, d\dot{x}^1\wedge d\dot{x}^2\wedge d\dot{x}^3
\end{equation}
where $|\dot{\bm{x}}|^2 = (\dot{x}^1)^2 + (\dot{x}^2)^2 +
(\dot{x}^3)^2$ and $\simeq$ denotes equality under restriction to
$\mathcal{E}$ by pull-back. The above are specialised to
inertial coordinates $(x^a)$ on Minkowski spacetime; their form in
a general coordinate system may be found in~\cite{burton_noble_wen:2009, ehlers:1971}.

We are interested in the evolution of a plasma over
timescales during which the motion of the ions is negligible in
comparison with the motion of the electrons. We assume that the ions are at
rest and distributed homogeneously in the laboratory frame. Their
worldlines are trajectories of the vector field $N_\text{ion} =
n_\text{ion}\partial/\partial x^0$ on $\mathcal{M}$ where
$n_\text{ion}$ is the ion number density (a positive definite constant) in the laboratory
frame. The Maxwell equations are
\begin{align}
&dF = 0,\\
\label{maxwell}
&d\star F = - q \star \widetilde{N} + q \star \widetilde{N_{\text{ion}}}
\end{align}
where $F =\frac{1}{2}F_{ab}\, dx^a\wedge dx^b$ is the electromagnetic
$2$-form and $q$ is the
charge on the electron (with $q<0$). The Hodge map $\star$ is induced
from (\ref{metric}) and the volume $4$-form $\star 1$,
\begin{equation}
\star 1 = dx^0\wedge dx^1\wedge dx^2\wedge dx^3,
\end{equation}
on ${\cal M}$. The $1$-forms $\widetilde{N}$,
$\widetilde{N_{\text{ion}}}$ are the metric duals of the vector fields
$N$, $N_{\text{ion}}$ respectively,
i.e. the $1$-form $\widetilde{Y}$ satisfies $\widetilde{Y}(Z)=g(Y,Z)$ for all
vector fields $Z$ on ${\cal M}$.

The scalar field $f$ satisfies the Vlasov equation, which may be written
\begin{equation}
\label{component_vlasov}
\dot{x}^a\bigg(\frac{\partial f}{\partial x^a} - \frac{q}{m}
F^b{ }_a^{\bm{V}}\frac{\partial f}{\partial\dot{x}^b} \bigg) \simeq 0
\end{equation}
on ${\cal E}$, where $F^b{ }_a^{\bm{V}}$ is the vertical lift of $F^b{ }_a =
\eta^{bc}F_{ca}$ from $\mathcal{M}$ to $T\mathcal{M}$,
\begin{equation}
F^b{ }_a^{\bm{V}}(x,\dot{x}) = F^b{ }_a(x),
\end{equation}
and $m$ is the electron rest mass.

The equations of motion for a waterbag distribution are readily
motivated via a global expression of the local Vlasov equation
(\ref{component_vlasov}).
Introduce the Liouville vector field $L$,
\begin{equation}
\label{definition_L}
L = \dot{x}^a\bigg(\frac{\partial}{\partial x^a} - \frac{q}{m}
F^b{ }_a^{\bm{V}}\frac{\partial}{\partial\dot{x}^b}\bigg),
\end{equation}
on $T\mathcal{M}$ and the $6$-form $\omega$,
\begin{equation}
\label{definition_omega}
\omega = \iota_L\,\iota_X (\star 1^{\bm{V}}\wedge \# 1)
\end{equation}
where the $4$-form $\star 1^{\bm{V}}$,
\begin{equation}
\star 1^{\bm{V}} = dx^0 \wedge dx^1 \wedge dx^2 \wedge dx^3,
\end{equation}
is the vertical lift of the spacetime volume $4$-form $\star 1$ from
$\mathcal{M}$ to $T\mathcal{M}$. It can be shown
\begin{equation}
d\omega \simeq 0
\end{equation}
and the Vlasov equation (\ref{component_vlasov}) can be written
\begin{equation}
d(f\omega) \simeq 0.
\end{equation}

Thus, it follows
\begin{equation}
\int_{\mathcal{B}} d(f\omega) = 0
\end{equation}
where $\mathcal{B}$ is a $7$-dimensional region in
$\mathcal{E}$ and using Stokes' theorem on forms (see,
for example,~\cite{burton:2003, benn:1987}) we obtain
\begin{equation}
\label{stokes_on_f_omega}
\int_{\mathcal{\partial B}} f\omega = 0
\end{equation}
where $\partial\mathcal{B}$ is the boundary of
$\mathcal{B}$.
\subsection{Waterbag distributions}
We consider distributions for which $f=\alpha$ is a positive constant inside a
$7$-dimensional
region $\mathcal{U}\subset\mathcal{E}$ and $f=0$ outside. In
particular, we consider $\mathcal{U}$ to be the union over each point
$x\in\mathcal{M}$ of a domain $\mathcal{W}_x$ whose boundary
$\partial\mathcal{W}_x$ in $\mathcal{E}$ is
topologically equivalent to the $2$-sphere.
Such piecewise constant distributions are called `waterbags'.

Choosing $\mathcal{B}$ in (\ref{stokes_on_f_omega}) to be a
small $7$-dimensional 
`pill-box' that intersects $\partial\mathcal{W}_x$ and evaluating
the integral in the limit as the `height' of $\mathcal{B}$ tends to zero, we
recover a jump condition on $f\omega$ that leads to
\begin{equation}
\label{jump_condition}
d\lambda\wedge \omega \simeq 0\,\,\,\text{at $\lambda=0$}
\end{equation}
where $\lambda=0$ is the union over $x$ of the boundaries
$\partial\mathcal{W}_x$. If $\lambda=0$ is the image of the embedding map $\Sigma$,
\begin{eqnarray}
\notag \Sigma : \mathcal{M} \times S^2 &\rightarrow& \mathcal{E} \\
(x, \xi) &\mapsto& (x, \dot{x} = V_\xi (x)), 
\end{eqnarray}
where $\xi\in S^2$ has coordinates $(\xi^1, \xi^2)$, then it
follows~\cite{burton_noble_wen:2009} from
(\ref{definition_X}, \ref{definition_L}, \ref{definition_omega}) that
(\ref{jump_condition}) is equivalent to
\begin{equation}
 \big( \nabla_{V_\xi} \widetilde{V_\xi} - \frac{q}{m} \iota_{V_\xi} F \big) \wedge
\Omega_\xi=0.  \label{Lorentz_jump}
\end{equation}
Here, $V_\xi$ and $\Omega_\xi$ are families of vector fields and 2-forms on
$\mathcal{M}$ respectively, where
\begin{align}
&V_\xi = V^a_\xi \frac{\partial}{\partial x^a},\\
\label{definition_Omega_xi}
&\Omega_\xi= \frac{\partial V^a_\xi}{\partial \xi^1} dx_a \wedge
\frac{\partial V^b_\xi}{\partial \xi^2} dx_b.
\end{align}
with $dx_a = \eta_{ab} dx^b$.  
Note that since the image of $\Sigma$ lies in $\mathcal{E}$ it
follows that, for each $\xi \in S^2$, $V_\xi$ is timelike, unit normalized and
future-directed:
\begin{align}
 \label{norm}
g(V_\xi,V_\xi) = -1,\qquad g(V_\xi, \frac{\partial}{\partial x^0}) <0.
\end{align}
We adopt (\ref{Lorentz_jump}) as the equation of motion for the
waterbag boundary
$\partial\mathcal{W}_x$. 

It may be shown that a particular class of solutions to
(\ref{Lorentz_jump}) satisfies
\begin{equation}
\label{solved_Lorentz}
F = \frac{m}{q} d \widetilde{V}_\xi
\end{equation}
and using (\ref{maxwell}) we obtain the field equation
\begin{equation}
\label{maxwell_2form_eliminated}
d\star d \widetilde{V}_\xi = - \frac{q^2}{m}(\star\widetilde{N} -
\star\widetilde{N_\text{ion}})
\end{equation}
on $\mathcal{M}$ with the condition that $d\widetilde{V}_\xi$ is independent of
$\xi$. For simplicity, we have neglected the direct contribution of
the driver (laser pulse or particle bunch) to the total electromagnetic
field in (\ref{solved_Lorentz}). 

For a discussion of solutions to (\ref{Lorentz_jump}) that do not satisfy
(\ref{solved_Lorentz}) see~\cite{burton_noble_wen:2009}.
\section{{ Electrostatic oscillations on 2-dimensional spacetime }}
\label{section:2d_discussion}
Before analysing (\ref{norm}, \ref{maxwell_2form_eliminated}) further
it is useful to briefly discuss their analogue on 2-dimensional
spacetime for facilitating comparison with the approach adopted
in~\cite{katsouleas:1988, mori:1990}.
\begin{figure}
\begin{center}
\scalebox{1.0}{\includegraphics{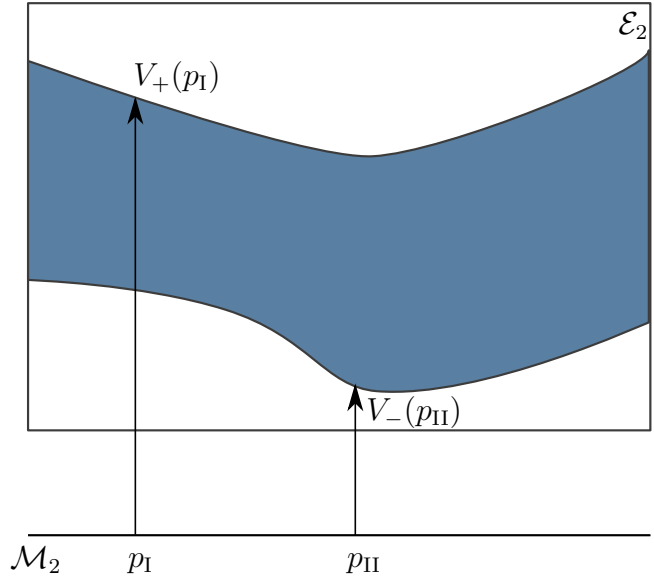}}
\caption{\label{fig:1D_waterbag} Illustration of a waterbag distribution over
  2-dimensional spacetime ${\cal M}_2$ with $p_\text{I} =
  (t_\text{I},z_\text{I})$ and $p_\text{II} =
  (t_\text{II},z_\text{II})$. The shaded region is the
  interior of the waterbag (where $f$ is non-zero), and the $2$-vector
  fields $\{V_+,V_-\}$ determine the boundary of the waterbag.}
\end{center}
\end{figure}
Although formulated on 4-dimensional spacetime, equations
(\ref{norm}, \ref{maxwell_2form_eliminated}) have a similar structure
for any number of dimensions. In particular, we now consider
2-dimensional Minkowski spacetime $(\mathcal{M}_2,g)$
\begin{align}
&g = - dt \otimes dt + dz \otimes dz,\\
&\star 1 = dt \wedge dz
\end{align}
where $(t,z)$\footnote{We use $(t,z)$ rather than $(x^a)$ to
distinguish coordinates on 2- and 4-dimensional spacetimes.} is a
Cartesian coordinate system in the laboratory inertial frame. An
induced coordinate system on $T\mathcal{M}_2$ is
$(t,z,\dot{t},\dot{z})$ and
the $2$-form $\# 1$ and dilation vector field $X$
over $T{\cal M}_2$ are
\begin{align}
&\# 1 = d\dot{t}\wedge d\dot{z},\\
&X = \dot{t}\frac{\partial}{\partial\dot{t}} + \dot{z}\frac{\partial}{\partial\dot{z}}.
\end{align}
Furthermore, $\xi$ is now an element of the
$0$-sphere $\{+,-\}$ and $\Omega_\xi = 1$ is a constant
$0$-form. Thus, the analogue to (\ref{Lorentz_jump}) is
\begin{align}
\label{Lorentz_jump2d_plus}
&\nabla_{V_+}
  \widetilde{V_+}
- \frac{q}{m} \iota_{V_+} F = 0,\\
\label{Lorentz_jump2d_minus}
&\nabla_{V_-}
  \widetilde{V_-}
- \frac{q}{m} \iota_{V_-} F = 0,
\end{align}
where $\{V_+,V_-\}$ satisfy the conditions
\begin{align}
\label{norm2d+}
&g(V_+,V_+) = -1,\qquad
g(V_+, \frac{\partial}{\partial t}) <0,\\
\label{norm2d-}
&g(V_-,V_-) = -1,\qquad
g(V_-, \frac{\partial}{\partial t}) <0,
\end{align}
and the only non-trivial Maxwell equation for the $2$-form $F$ is 
\begin{equation}
\label{max2d}
d \star F = - q \star \widetilde{N} + q \star \widetilde{N_{\text{ion}}}
\end{equation}
where $N_{\text{ion}} = n_{\text{ion}}\partial/\partial t$ is the ion number
$2$-current and $F = E dt\wedge dz$ where $E$ is the electric field along
the $z$-axis.

On the unit hyperbola bundle $\mathcal{E}_2$, $\dot{t}=
\sqrt{1+\dot{z}^2}$ and the components of the electron number
$2$-current $N=N^t \partial/\partial t + N^z \partial/\partial z$
corresponding to
(\ref{component_number_current_forms}) are
\begin{align}
&\nonumber N^t = \int_\mathbb{R} f(t,z,\dot{t},\dot{z})
\,d\dot{z} = \alpha \Big( Y_+- Y_- \Big), \\
&N^z = \int_\mathbb{R} \frac{\dot{z}}{\sqrt{1+\dot{z}^2}} f(t,z,\dot{t},\dot{z})
\,d\dot{z}= \alpha \Big( \sqrt{1+Y^2_+}- \sqrt{1+Y^2_-} \Big),
\end{align}
where
\begin{equation}
f =
\begin{cases}
\alpha,\qquad Y_-\le \dot{z} \le Y_+,\\
0, \qquad \dot{z} < Y_-\text{ or } \dot{z} > Y_+
\end{cases}
\end{equation}
with $\alpha$ a positive non-zero constant and $\{Y_+, Y_-\}$ $0$-forms
on ${\cal M}_2$.  The $2$-velocity fields
$\{V_+,V_-\}$ satisfy
\begin{equation}
V_\pm = \sqrt{1+Y^2_{\pm}}\, \frac{\partial}{\partial t} + Y_{\pm} \,
\frac{\partial}{\partial z}
\end{equation}
and it follows
\begin{equation}
\label{2dcurrent_def}
\widetilde{N}= \alpha {\star \big( \widetilde{V_+}- \widetilde{V_-} \big)}.
\end{equation}
See figure~\ref{fig:1D_waterbag}.

Unlike their $4$-dimensional analogue, which may include transverse
electromagnetic fields,
(\ref{Lorentz_jump2d_plus},\ref{Lorentz_jump2d_minus}) are
\emph{uniquely}\footnote{Proper
incorporation of transverse fields requires at
least $2$ spatial dimensions.} solved by  
\begin{equation}
d\widetilde{V_\pm} = \frac{q}{m} F \label{Lorentz2d}
\end{equation}
and using (\ref{max2d}) it follows
\begin{equation}
\label{maxwell_2form_eliminated2d}
d \star d \widetilde{V_\pm} = - \frac{q^2}{m} (\star \widetilde{N} - \star
\widetilde{N_{\text{ion}}})
\end{equation}
subject to the condition $d\widetilde{V_+} = d\widetilde{V_-}$.

Alternatively, one may follow the approach adopted in
\cite{katsouleas:1988} by casting the above as a warm fluid.
The type $(0,2)$ stress-energy-momentum tensor
$\mathcal{T}_\text{fluid}$ of the electron fluid is
\begin{equation}
\label{stress_2d_fluid}
\mathcal{T}_\text{fluid} = \bigg(m\int_\mathbb{R} \frac{\dot{z}^\mu \dot{z}^\nu
f}{\sqrt{1+\dot{z}^2}} d\dot{z}
\bigg)\frac{\partial}{\partial
  z^\mu}\otimes\frac{\partial}{\partial z^\nu}
\end{equation}
where Greek indices $\mu,\nu$ run over $0,1$ and $z^0=t,\, z^1=z,\,
\dot{z}^0=\sqrt{1+\dot{z}^2},\, \dot{z}^1=\dot{z}$. It
can be shown that $\mathcal{T}_\text{fluid}$ induced by the above waterbag
distribution can be expressed entirely in terms of the proper number
density $n$ of the electron fluid, the electron fluid's bulk
$2$-velocity $U$ and the spacetime metric:
\begin{equation}
\mathcal{T}_\text{fluid} = (\rho + p) U\otimes U + p\,\widetilde{g}
\end{equation}
where $\widetilde{g}$ is the inverse metric tensor
\begin{equation}
\tilde{g} = -\frac{\partial}{\partial t} \otimes
 \frac{\partial}{\partial t} + \frac{\partial}{\partial z} \otimes \frac{\partial}{\partial z}
\end{equation}
and
\begin{align}
\label{2d_U_def}
&U = \frac{1}{\sqrt{-g(Z,Z)}} Z,\qquad Z = \frac{1}{2}(V_+ + V_-),\\
\label{2d_n_def}
&N = nU,\qquad n = \sqrt{-g(N,N)},
\end{align}
with the equations of state
\begin{align}
\label{eqns_state2da}
&\rho =
  m\alpha\bigg[\frac{n}{2\alpha}\sqrt{1+\bigg(\frac{n}{2\alpha}\bigg)^2}
  + \text{sinh}^{-1}\bigg(\frac{n}{2\alpha}\bigg)\bigg],\\
\label{eqns_state2db}
&p = m\alpha\bigg[\frac{n}{2\alpha}\sqrt{1+\bigg(\frac{n}{2\alpha}\bigg)^2}
  - \text{sinh}^{-1}\bigg(\frac{n}{2\alpha}\bigg)\bigg].
\end{align} 
The equation of motion of the electron fluid,
\begin{align}
\label{warm_fluid2d}
(\rho + p) \nabla_U \widetilde{U} = qn \iota_U F - \iota_U
  ( dp \wedge \widetilde{U} ),
\end{align}
follows from the zero divergence of the sum of
$\mathcal{T}_\text{fluid}$ and the Maxwell stress-energy-momentum
tensor where
\begin{equation}
\qquad g(U,U) = -1,\qquad g(U,\frac{\partial}{\partial t}) < 0.
\end{equation}

It should be stressed that the warm fluid model
(\ref{eqns_state2da}, \ref{eqns_state2db}, \ref{warm_fluid2d}) is
equivalent to (\ref{Lorentz_jump2d_plus}, \ref{Lorentz_jump2d_minus},
\ref{norm2d+}, \ref{norm2d-}).
Thus, (\ref{Lorentz_jump2d_plus}, \ref{Lorentz_jump2d_minus},
\ref{norm2d+}, \ref{norm2d-}) may be replaced by an equivalent field
theory expressed in terms of a finite set of moments of $f$ on
$2$-dimensional spacetime. However, the situation is more complicated
for waterbags over $4$-dimensional spacetime where second, and higher, order
moments of $f$ in $\dot{x}$ are not, in general, easily expressible in terms of zeroth and
first order moments of $f$. In general, the moment hierarchy does not
automatically close. 

We will now obtain a non-linear ordinary differential equation
describing 1-dimensional
electrostatic oscillations and determine an expression for the wave-breaking
limit of this model. Derivation of wave-breaking limits starting
from $\mathcal{T}_\text{fluid}$ and the equations of state
(\ref{eqns_state2da}, \ref{eqns_state2db}) may be found in
\cite{katsouleas:1988, mori:1990}. However, we will work directly with
(\ref{norm2d+}, \ref{norm2d-}, \ref{2dcurrent_def},
\ref{maxwell_2form_eliminated2d})
to facilitate comparison with our model on $4$-dimensional spacetime.

Let all field components with respect to the laboratory frame $(dt,dz)$ be functions
of $\zeta = z - vt$ only
(the `quasi-static assumption'), where $0<v<1$, and let $(e^1,e^2)$ be the basis
\begin{equation}
e^1= vdz-dt, \qquad e^2=dz-vdt.
\end{equation}
The coframe $(\gamma e^1,\gamma e^2)$ is an orthonormal
basis adapted to observers moving at velocity $v$ along $z$ (i.e observers in the
`wave frame') where
$\gamma= (1-v^2)^{-1/2}$ is the Lorentz factor of such observers
relative to the laboratory. So, $\gamma e^2(N_\text{ion}) = -
\gamma n_\text{ion} v$ is the component of the ion number $1$-current
in the wave frame.

In the basis $(e^1,e^2)$, $\widetilde{V_\pm}$ can be decomposed as
\begin{equation}
\widetilde{V_\pm}= \big( \mu(\zeta) + A_\pm \big) e^1 + \psi_\pm
(\zeta) e^2   \label{decomp2d}
\end{equation}
where $\{A_+,A_-\}$ are constant.
Note that this is the most general decomposition compatible with
equation (\ref{Lorentz2d}) and the quasi-static assumption. 

Solving (\ref{norm2d+}, \ref{norm2d-}) for $\psi^2_\pm$ gives
\begin{equation}
\psi^2_\pm = (\mu + A_\pm)^2- \gamma^2
\end{equation}
and additional physical information is needed to fix
the sign of $\psi_\pm$. Here, we demand that all electrons
described by the waterbag are
travelling slower than the wave so $\psi_\pm = -\sqrt{(\mu + A_\pm)^2- \gamma^2}$
and (\ref{decomp2d}) is 
\begin{align}
\widetilde{V_\pm} = &\big( \mu + A_\pm \big) e^1
- \Big( (\mu +A_\pm)^2- \gamma^2 \Big)^{1/2}  e^2.  
\end{align}

Substituting (\ref{decomp2d}) into equation (\ref{Lorentz2d}) yields
\begin{equation}
\label{electric2d}
E= \frac{1}{\gamma^2} \frac{m}{q} \frac{d\mu}{d\zeta},
\end{equation}
and equation (\ref{maxwell_2form_eliminated2d}) yields the nonlinear oscillator
equation
\begin{equation}
\label{osc2d}
\frac{1}{\gamma^2} \frac{d^2 \mu}{d\zeta^2} = -\frac{q^2}{m} \gamma^2 n_\text{ion} -
\frac{q^2}{m} \alpha 
\bigg[ \sqrt{(\mu + A_+)^2 -\gamma^2} -\sqrt{(\mu + A_-)^2 -\gamma^2} \bigg]
\end{equation}
with the algebraic constraint
\begin{equation}
\label{alg_const2d}
A_+ -A_- = -\frac{n_\text{ion} \gamma^2 v}{\alpha} < 0.
\end{equation}
\subsection{Electrostatic wave-breaking}
In the wave frame the relativistic energies of the two ends of the waterbag
are $m(\mu + A_+)/\gamma$ and $m(\mu + A_-)/\gamma$ respectively, and
since $m(\mu + A_+)/\gamma \ge m$ it follows $\mu + A_\pm \ge
\gamma$. Using (\ref{alg_const2d}), $\mu + A_- > \mu + A_+$ and hence
$\mu+A_+\ge\gamma$ implies $\mu+A_->\gamma$. Thus, $\mu+A_\pm\ge\gamma$
may be reduced to $\mu \ge \mu_\text{wb}$ where
\begin{equation}
\label{2d_muwb_def}
\mu_\text{wb} = -A_+ + \gamma.
\end{equation}
Alternatively, one may arrive at the same conclusion by inspecting the
right-hand side of (\ref{osc2d}) and using $\mu + A_\pm > 0$ (which
follows because $V_+$ and $V_-$ are future-pointing). 
Thus, there is an upper bound on the amplitude of oscillatory solutions to (\ref{osc2d}), 
which leads to an upper bound $E_\text{max}$ on the electric field $E$ (the
`wave-breaking limit' of this model).

During an oscillation $E$
vanishes when $d\mu/d\zeta$ vanishes and $|E|$ is at a maximum when
$|d\mu/d\zeta|$ is at a maximum (see (\ref{electric2d}) and note $q<0$). A maximum of $|d\mu/d\zeta|$ occurs at
values $\zeta_0$ of $\zeta$ where $\mu(\zeta_0)$ equals the oscillator
equilibrium $\mu_\text{eq}$. Furthermore, for the
maximum amplitude oscillation $d\mu/d\zeta$ vanishes when
$\mu=\mu_\text{wb}$. An upper bound $E_\text{max}$ on the magnitude
$|E|$ of the electric field is obtained by evaluating the first integral of
(\ref{osc2d}) between $\mu=\mu_\text{wb}$ and $\mu=\mu_\text{eq}$. 

Without loss of generality, we can choose the split between $\mu$ and
 $A_\pm$ such that
\begin{equation}
\label{2d_a_def}
A_+ = -A_- = -a,\qquad a = \frac{n_\text{ion}\gamma^2 v}{2\alpha}.
\end{equation}
Using (\ref{electric2d}, \ref{osc2d}) it follows
\begin{equation}
\label{2d_Emax2}
E^2_\text{max} = 2mn_\text{ion}\bigg[-\mu_\text{eq}+\mu_\text{wb} +
  \frac{1}{2}\frac{v}{a}\int\limits^{\mu_\text{eq}}_{\mu_\text{wb}}\bigg(\sqrt{[\mu+a]^2-\gamma^2}
  - \sqrt{[\mu-a]^2-\gamma^2}\bigg)d\mu\bigg]
\end{equation}
where $\mu_\text{eq}$ is the equilibrium solution to (\ref{osc2d}),
which satisfies
\begin{equation}
\label{2d_mueq_def}
\frac{2a}{v} = \sqrt{(\mu_\text{eq}+a)^2-\gamma^2} - \sqrt{(\mu_\text{eq}-a)^2-\gamma^2}.
\end{equation}

The constant $a$ is fixed in terms of an effective temperature
$T_\text{eq}$ associated with the oscillator equilibrium
$\mu_\text{eq}$. Noting that $n=n_\text{ion}$ when the waterbag is in
its equilibrium state ($\mu=\mu_\text{eq}$), and assuming
$n_\text{ion} \ll 2\alpha$ in (\ref{eqns_state2db}) it follows
\begin{equation}
p_\text{eq} \approx \frac{m n^3_\text{ion}}{3(2\alpha)^2}.
\end{equation}
Introducing $T_\text{eq}$ via $p_\text{eq} = n_\text{ion} k_B
T_\text{eq}$, where $k_B$ is Boltzmann's constant, we find
\begin{equation}
\label{2d_Teq_def}
\frac{n_\text{ion}}{2\alpha} \approx \sqrt{\frac{3 k_B T_\text{eq}}{m}},\qquad a
  \approx \gamma^2 v\sqrt{\frac{3 k_B T_\text{eq}}{m}}
\end{equation}
where (\ref{2d_a_def}) has been used. Hence, $n_\text{ion}\ll 2\alpha$
means that the thermal energy of the electron fluid in the oscillator
equilibrium state is much less than the rest mass-energy of the
electron.

The wave-breaking limit $E_\text{max}$ can be readily analysed for $\gamma\gg 1$
via asymptotic approximation in a small parameter $\varepsilon$,
\begin{equation}
\label{2d_varepsilon_def}
\varepsilon = \frac{\gamma}{a} = \frac{2\alpha}{n_\text{ion}v}\frac{1}{\gamma},
\end{equation}
where (\ref{2d_a_def}) has been used.
Employing (\ref{2d_muwb_def}, \ref{2d_Emax2}, \ref{2d_mueq_def}) it
follows
\begin{align}
\label{2d_Emax2_hatted}
&E^2_\text{max} = 2mn_\text{ion}a\bigg[-\hat{\mu}_\text{eq}+\hat{\mu}_\text{wb} +
  \frac{1}{2}v\int\limits^{\hat{\mu}_\text{eq}}_{\hat{\mu}_\text{wb}}\bigg(\sqrt{[\hat{\mu}+1]^2-\varepsilon^2}
- \sqrt{[\hat{\mu}-1]^2-\varepsilon^2}\bigg)d\hat{\mu}\bigg],\\
\label{2d_mueq_def_hatted}
&\frac{1}{2}v\bigg(\sqrt{[\hat{\mu}_\text{eq} + 1]^2 - \varepsilon^2}
- \sqrt{[\hat{\mu}_\text{eq} - 1]^2 - \varepsilon^2}\bigg) = 1,\\
\label{2d_muwb_def_hatted}
&\hat{\mu}_\text{wb} = 1 + \varepsilon 
\end{align}
where, $\hat{\mu}=\mu/a$. To proceed further, we express $v$ in
(\ref{2d_Emax2_hatted}, \ref{2d_mueq_def_hatted}) as a function of
$\varepsilon$ and a parameter $b$ that characterizes the effective
temperature of the oscillator equilibrium distribution. Using
(\ref{2d_a_def}, \ref{2d_varepsilon_def}) it follows
\begin{equation}
\label{2d_v_varepsilon_def}
v = \frac{1}{\sqrt{1+\varepsilon^2 b^2}},
\end{equation}
where $b$ is
\begin{equation}
\label{2d_b_def}
b = \frac{n_\text{ion}}{2\alpha}.
\end{equation}

The dominant $\varepsilon$ dependence (as $\varepsilon\rightarrow 0$
with $b$ held fixed)
of $E^2_\text{max}$ arises from the second term in the integrand in
(\ref{2d_Emax2_hatted}) and
may be extracted by expanding the integrand with respect to $\varepsilon$ and
integrating each summand over $\hat{\mu}$. Since, for $\nu > \varepsilon > 0$, 
\begin{equation}
\sqrt{\nu^2-\varepsilon^2} = \nu -
\frac{1}{2}\frac{\varepsilon^2}{\nu} +
  \sum\limits^\infty_{n=2} c_n \frac{\varepsilon^{2n}}{\nu^{2n-1}}
\end{equation}
where $c_n$ are constants, and inspection of (\ref{2d_mueq_def_hatted}) reveals
\begin{equation}
\hat{\mu}_\text{eq} = h(\varepsilon^2) = h(0) + h'(0)\varepsilon^2 +
{\cal O}(\varepsilon^4)\qquad(\varepsilon\rightarrow 0),
\end{equation}
we find, for $h(0)\gg 1$,
\begin{align}
\notag
\int\limits^{\hat{\mu}_\text{eq}}_{\hat{\mu}_\text{wb}} \sqrt{[\hat{\mu}-1]^2 -
  \varepsilon^2} d\hat{\mu} &= \bigg(\frac{1}{2}[\hat{\mu}-1]^2 -
  \frac{1}{2}\varepsilon^2\ln(\hat{\mu}-1)\bigg)\bigg|^{\hat{\mu}_\text{eq}}_{\hat{\mu}_\text{wb}}
+ \sum\limits^\infty_{n=2} c_n
  \frac{1}{2-2n}\bigg(\frac{\varepsilon^{2n}}{[\hat{\mu}_\text{eq}-1]^{2n-2}}
  - \varepsilon^2\bigg)\\
&= \frac{1}{2}[\hat{\mu}-1]^2\bigg|^{\hat{\mu}_\text{eq}}_{\hat{\mu}_\text{wb}}
  + \frac{1}{2}\varepsilon^2\ln(\varepsilon) + {\cal
  O}(\varepsilon^2)\qquad (\varepsilon\rightarrow 0)
\end{align}
where (\ref{2d_muwb_def_hatted}) has been used.
Furthermore, it follows from (\ref{2d_mueq_def_hatted}) that an
asymptotic approximation for $h(0)$ in small $b$ leads to
\begin{equation}
h(0) = \frac{1}{b} + {\cal O}(1)\qquad (b\rightarrow 0).
\end{equation}
Thus, (\ref{2d_Emax2_hatted}) yields
\begin{equation}
\frac{E_\text{max}^2}{a} \approx -\frac{1}{2}m
n_\text{ion}\varepsilon^2\ln(\varepsilon)
\end{equation}
for $\varepsilon, b \ll 1$. Introducing the effective temperature
$T_\text{eq}$ using (\ref{2d_a_def}, \ref{2d_Teq_def},
\ref{2d_varepsilon_def}) and noting $v\approx 1$, we obtain
\begin{equation}
\label{2d_Emax2_final}
E_\text{max}^2 \approx \frac{1}{2}\frac{m^2 c^2 \omega_p^2}{q^2}
    \sqrt{\frac{mc^2}{3 k_B T_\text{eq}}}
\ln\bigg(\gamma\sqrt{\frac{3 k_B
    T_\text{eq}}{mc^2}}\bigg),\qquad \varepsilon, b \ll 1 
\end{equation}
where $\omega_p=\sqrt{n_\text{ion}q^2/(m\varepsilon_0)}$ is the plasma
frequency and the speed of light $c$ and permittivity of the
vacuum $\varepsilon_0$ have been restored. Equation
(\ref{2d_Emax2_final}) was obtained as a lower bound on
$E_\text{max}^2$ in~\cite{trines:2006}.
\section{Longitudinal electrostatic oscillations on 4-dimensional spacetime}
We now consider longitudinal electrostatic waves on 4-dimensional spacetime by closely
following the above description on 2-dimensional spacetime.  

As before, we adopt the `quasi-static assumption'. We seek a waterbag
$\mathcal{W}_x$ axisymmetric
about $\dot{x}^3$ whose pointwise dependence in Minkowski spacetime
$\mathcal{M}$ is on the wave's phase $\zeta = x^3 - v x^0$ only, where
$0<v<1$. As before, the following results are applicable only if the
longitudinal component of $V_\xi$ in the wave frame is negative
(no electron described by $\mathcal{W}_x$ is moving faster along $x^3$
than the wave).

Decompose $\widetilde{V}_\xi$ in the wave frame as
\begin{equation}
\label{V_ansatz}
\widetilde{V}_\xi = [\mu(\zeta) + A(\xi^1)]\, e^1 + \psi(\xi^1,\zeta)\, e^2
\,\,+ R\sin(\xi^1)\cos(\xi^2)dx^1 + R\sin(\xi^1)\sin(\xi^2)dx^2
\end{equation}
for $0 < \xi^1 < \pi$, $0 \le \xi^2 < 2\pi$ 
where $R>0$ is constant and
\begin{equation}
\label{coframe}
e^1 = v dx^3 - dx^0,\qquad e^2 = dx^3 - v dx^0.
\end{equation}
Here, $(\gamma e^1, \gamma e^2, dx^1, dx^2)$ is an orthonormal basis
adapted to the wave frame, with $\gamma = 1/\sqrt{1-v^2}$.
In the wave frame the relativistic energy of $P_\xi = m V_\xi$ is $m(\mu +
A(\xi^1))/\gamma$ and it follows that $\mu+A(\xi^1) > 0$. Furthermore, using
(\ref{norm}, \ref{V_ansatz}) it follows
\begin{equation}
\label{psi}
\psi = -\sqrt{[\mu + A(\xi^1)]^2 - \gamma^2[1 + R^2 \sin^2(\xi^1)]},
\end{equation}
where the negative square root is chosen because no electron is moving
faster along $x^3$ than the wave, and we obtain
\begin{equation}
\mu \ge - A(\xi^1) +
\gamma\sqrt{1+R^2\sin^2(\xi^1)}.
\end{equation} 
 
Substituting (\ref{V_ansatz}) into equation (\ref{solved_Lorentz}) leads to 
\begin{equation}
\label{F_dmu}
F= \frac{m}{q} \frac{d \mu}{d\zeta} e^2 \wedge e^1,
\end{equation}
and (\ref{component_number_current_forms},
\ref{maxwell_2form_eliminated}, \ref{V_ansatz}, \ref{psi}) yield
\begin{equation}
\frac{1}{\gamma^2}\frac{d^2\mu}{d\zeta^2} = -
  \frac{q^2}{m}n_{\text{ion}}\gamma^2
- \frac{q^2}{m}2\pi R^2 \alpha \int\limits^\pi_0 \bigg([\mu +
    A(\xi^1)]^2
\,\,- \gamma^2[1 + R^2
    \sin^2(\xi^1)]\bigg)^{1/2}\sin(\xi^1)\,\cos(\xi^1)\, d\xi^1
  \label{ODE_mu}
\end{equation}
(c.f. equation (\ref{osc2d})) and
\begin{equation}
\label{norm_A}
2\pi R^2 \int\limits^\pi_0
A(\xi^1)\,\sin(\xi^1)\,\cos(\xi^1)\,d\xi^1 = - \frac{n_\text{ion}\gamma^2\,v}{\alpha}
\end{equation}
(c.f. equation (\ref{alg_const2d})) where $\alpha>0$ is the value of $f$ inside
$\mathcal{W}_x$.

The form of the 2nd order autonomous
non-linear ordinary differential equation (\ref{ODE_mu}) for $\mu$ is fixed by
specifying the generator $A(\xi^1)$ of $\partial\mathcal{W}_x$ subject to the
normalization condition (\ref{norm_A}).
\subsection{Electrostatic wave-breaking}
The form of the integrand in (\ref{ODE_mu}) ensures that the magnitude
of oscillatory solutions to (\ref{ODE_mu}) cannot be arbitrarily
large. For our model, the wave-breaking value $\mu_{\text{wb}}$
is the largest $\mu$ for which the argument of the square root in
(\ref{ODE_mu}) vanishes,
\begin{equation}
\mu_{\text{wb}} =
\text{max}\bigg\{-A(\xi^1) + \gamma\sqrt{1+R^2\sin^2(\xi^1)}
\,\bigg|\,0\le\xi^1\le\pi\bigg\},  \label{mu_wave-breaking}
\end{equation}
because $\mu<\mu_{\text{wb}}$ yields an imaginary integrand in
(\ref{ODE_mu}) for some
$\xi^1$.

The electric field has only one non-zero component $E$ (in the $x^3$
direction). Using $F=E\,dx^0\wedge dx^3$ and (\ref{coframe},
\ref{F_dmu}) it follows
\begin{equation}
\label{E_dmu}
E = \frac{m}{q} \frac{1}{\gamma^2} \frac{d\mu}{d\zeta}
\end{equation}
and the wave-breaking limit $E_{\text{max}}$ is obtained by evaluating
the first integral of (\ref{ODE_mu}) between $\mu_{\text{wb}}$ where
$E$ vanishes and the oscillator
equilibrium $\mu_{\text{eq}}$ of $\mu$ where $|E|$ is at a maximum. Using
(\ref{norm_A}) to eliminate $\alpha$ it follows that $\mu_{\text{eq}}$ satisfies
\begin{equation}
\frac{1}{v}\int\limits^\pi_0 A(\xi^1)\sin(\xi^1)\cos(\xi^1)\,d\xi^1
= \int\limits^\pi_0 \bigg([\mu_{\text{eq}} + A(\xi^1)]^2
- \gamma^2[1 + R^2
    \sin^2(\xi^1)]\bigg)^{1/2}
\sin(\xi^1)\cos(\xi^1)
    d\xi^1   \label{mu_equilibrium}
\end{equation}
with 
\begin{equation}
\label{A_negativity}
\int\limits^\pi_0 A(\xi^1)\sin(\xi^1)\cos(\xi^1)\,d\xi^1\, <\, 0
\end{equation}
since $\alpha, v >0$. Equation (\ref{ODE_mu}) yields the
maximum value $E_\text{max}$ of $E$,
\begin{align}
\notag E_{\text{max}}^2 = 2 m n_\text{ion}\Bigg[
-\mu_{\text{eq}} + \mu_{\text{wb}}
+ \, &\frac{v}{\int\limits^\pi_0
    A(\xi^{1\prime})\sin(\xi^{1\prime})\cos(\xi^{1\prime})d\xi^{1\prime}} \times \\
& \int\limits^{\mu_{\text{eq}}}_{\mu_{\text{wb}}}\int\limits^\pi_0
\bigg([\mu + A(\xi^1)]^2
- \gamma^2 [1 + R^2
    \sin^2(\xi^1)]\bigg)^{1/2}
\sin(\xi^1)\cos(\xi^1)
    d\xi^1\,d\mu\Bigg].   \label{E_max}
\end{align}

To proceed further we need to choose the generator $A(\xi^1)$ of the waterbag
distribution. It turns out that even the simple choice
\begin{equation}
A(\xi^1) = -a\cos(\xi^1)
\end{equation}
for $A(\xi^1)$, where $a$ is a positive constant, leads to a
wave-breaking limit $E_\text{max}$ with interesting behaviour, as we
now show.

Using (\ref{E_max}) it follows
\hspace{-2em}
\begin{equation}
E_{\text{max}}^2 = 2 m n_\text{ion}\Bigg[
-\mu_{\text{eq}} + \mu_{\text{wb}}
+ \frac{3}{2}\frac{v}{a}
\int\limits^{\mu_{\text{eq}}}_{\mu_{\text{wb}}}\int\limits^1_{-1}
\bigg([\mu + a\chi]^2
- \gamma^2[1 + R^2 (1-\chi^2)]\bigg)^{1/2}\chi\,d\chi\,d\mu
\Bigg]   \label{E_max_example}
\end{equation}
where $\chi=-\cos(\xi^1)$, equation (\ref{mu_equilibrium}) yields
\begin{equation}
\frac{3}{2}\frac{v}{a}\int\limits^1_{-1} \bigg([\mu_{\text{eq}} +
  a\chi]^2
- \gamma^2[1 + R^2(1-\chi^2)]\bigg)^{1/2}\chi\,d\chi = 1  
\label{mu_equilibrium_example}
 \end{equation}
and equation (\ref{mu_wave-breaking}) may be written
\begin{equation}
\mu_{\text{wb}} =
\text{max}\bigg\{-a\chi + \gamma\sqrt{1+R^2(1-\chi^2)}
\,\bigg|\,-1\le\chi\le 1\bigg\}.   \label{mu_wave-breaking_max_example}
\end{equation}
\begin{figure}
\begin{center}
\scalebox{1.0}{\includegraphics{3D_waterbag_near_breaking_bowl.epsi}}
\qquad
\scalebox{1.0}{\includegraphics{3D_waterbag_near_breaking_bowl_wireframe.epsi}}
\caption{\label{fig:bowl} Two illustrations of the $4$-velocity dependence
  of a particular `bowl' waterbag. The axis of symmetry is aligned along
  $\dot{x}^3$. The maximum electric field amplitude is achieved during
  the oscillation in which the top of the waterbag (a circle) grazes the
  phase speed of the wave.} 
\end{center}
\end{figure}
\begin{figure}
\begin{center}
\scalebox{1.0}{\includegraphics{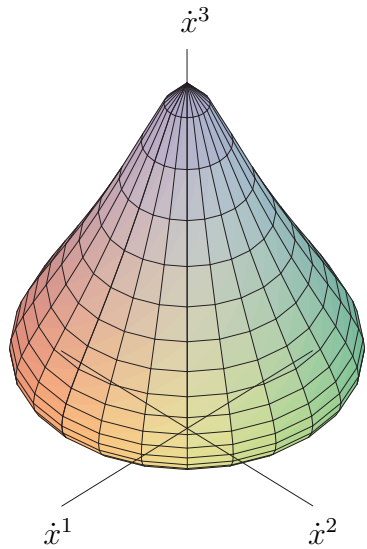}}
\caption{\label{fig:gourd} An illustration of the $4$-velocity dependence
  of a particular `gourd' waterbag. The axis of symmetry is aligned along
  $\dot{x}^3$. The  maximum electric field amplitude is achieved
  during the oscillation in which the tip of the waterbag grazes the
  phase speed of the wave.}
\end{center}
\end{figure}
Examination of (\ref{mu_wave-breaking_max_example}) reveals that two
classes of waterbag arise according to whether or not the function
\begin{equation}
\label{chi_map}
\chi \mapsto -a\chi + \gamma\sqrt{1+R^2(1-\chi^2)}
\end{equation}
has a turning point in the interval $[-1,1]$. Examples of the two classes
are shown in figures \ref{fig:bowl} and \ref{fig:gourd}. In each case,
the plasma wave breaks when the uppermost part of the distribution
achieves the phase velocity of the plasma wave (i.e. the
longitudinal component $\psi$ of $V_\xi$ in the wave frame
vanishes). Wave-breaking limits for the class in figure
\ref{fig:bowl} have been calculated
previously~\cite{burton_noble_wen:2009, burton_noble:2009} and here we
focus on waterbags of the type shown in figure~\ref{fig:gourd}.
\subsubsection{Calculation of the maximum electric field}
The parameters $a,R,\gamma$ are chosen to satisfy
\begin{equation}
\frac{a}{R}\sqrt{\frac{1+R^2}{a^2+\gamma^2 R^2}} > 1,
\end{equation}
ensuring that (\ref{chi_map}) does not have a turning point in the interval $[-1,1]$. Hence,
the wave breaks when the tip $\chi=-\cos(0)=-1$ of the waterbag achieves the phase velocity of the plasma
wave. Using (\ref{mu_wave-breaking_max_example}) it follows
\begin{equation}
\label{mu_wave-breaking_gourd}
\mu_{\text{wb}} = a + \gamma
\end{equation}
which is formally identical to the wave-breaking limit of $\mu$ for
the waterbag over $2$-dimensional spacetime. This is quite different
from the value of $\mu_{\text{wb}}$ for waterbags of the type shown
in figure~\ref{fig:bowl} (see~\cite{burton_noble_wen:2009, burton_noble:2009}).

Following a similar method to that used in section
\ref{section:2d_discussion}, we now evaluate (\ref{E_max_example}) for
$\gamma\gg 1$. Introducing $\hat{\mu}=\mu/a$ in
(\ref{E_max_example}, \ref{mu_equilibrium_example}, \ref{mu_wave-breaking_gourd})
leads to
\begin{align}
\label{E_max_gourd_hat}
&E_{\text{max}}^2 = 2 m n_\text{ion} a \Bigg[
-\hat{\mu}_{\text{eq}} + \hat{\mu}_{\text{wb}}
+ \frac{3}{2}v
\int\limits^{\hat{\mu}_{\text{eq}}}_{\hat{\mu}_{\text{wb}}}\int\limits^1_{-1}
\bigg([\hat{\mu} + \chi]^2
- \varepsilon^2[1 + R^2 (1-\chi^2)]\bigg)^{1/2}\chi\,d\hat{\mu}\,d\chi
\Bigg],\\
\label{mu_equilibrium_gourd_hat}
&\frac{3}{2}v\int\limits^1_{-1} \bigg([\hat{\mu}_{\text{eq}} +
  \chi]^2
- \varepsilon^2[1 + R^2(1-\chi^2)]\bigg)^{1/2}\chi\,d\chi = 1,\\
\label{mu_wave-breaking_gourd_hat}
&\hat{\mu}_{\text{wb}}
= 1 + \varepsilon,
 \end{align}
where, using (\ref{norm_A}),
\begin{align}
\label{def_a}
a &= \frac{3 n_\text{ion} \gamma^2 v}{4\pi R^2 \alpha}\\
\label{def_varepsilon}
\varepsilon &= \frac{\gamma}{a} = \frac{4\pi R^2 \alpha}{3
  n_{\text{ion}}\gamma v}.
\end{align}
Thus, it follows $\varepsilon\rightarrow 0$ as $v\rightarrow 1$ and we
determine an asymptotic approximation for $E_{\text{max}}$ in
$\varepsilon$ as $\varepsilon\rightarrow 0$.

Expansion in $\varepsilon$ of the integrand in (\ref{E_max_gourd_hat}) yields
\begin{equation}
\label{integrand_varpesilon_series}
\bigg([\hat{\mu} + \chi]^2
- \varepsilon^2[1 + R^2 (1-\chi^2)]\bigg)^{1/2} = \hat{\mu} + \chi -
\frac{1+R^2(1-\chi^2)}{2(\hat{\mu}+\chi)}\varepsilon^2
+ \sum\limits^\infty_{n=2} c_n
\frac{(1+R^2(1-\chi^2))^n}{(\hat{\mu}+\chi)^{2n-1}} \varepsilon^{2n}
\end{equation}
where the $c_n$ are numerical constants. Using
(\ref{integrand_varpesilon_series}), the integral over $\hat{\mu}$ in
(\ref{E_max_gourd_hat}) leads to a summand proportional to
\begin{equation}
\label{summand}
f_n = \int\limits^1_{-1}\bigg[\frac{(1+R^2(1-\chi^2))^n}{(\hat{\mu}_\text{eq}+\chi)^{2n-2}}
  -
  \frac{(1+R^2(1-\chi^2))^n}{(1+\varepsilon+\chi)^{2n-2}}\bigg]\chi\,
d\chi,\qquad n\ge 2
\end{equation}
where (\ref{mu_wave-breaking_gourd_hat}) has been used.

Inspection of (\ref{mu_equilibrium_gourd_hat}) suggests 
an approximation for $\hat{\mu}_\text{eq}(\varepsilon)$ of the form
\begin{equation}
\hat{\mu}_\text{eq}(\varepsilon) = h(\varepsilon^2) = h(0) +
h^\prime(0)\varepsilon^2 +
{\cal O}(\varepsilon^4)\qquad (\varepsilon\rightarrow 0).
\end{equation}
Using (\ref{def_varepsilon}) it follows
\begin{align}
\label{def_v_varepsilon}
&v = \frac{1}{\sqrt{1+\varepsilon^2 b^2}} = 1 -
\frac{1}{2}\varepsilon^2 b^2 + {\cal O}(\varepsilon^4)\qquad (\varepsilon\rightarrow 0),\\
\label{def_b}
&b = \frac{3 n_\text{ion}}{4\pi R^2\alpha} = \frac{a}{\gamma^2 v}
\end{align}
and (\ref{mu_equilibrium_gourd_hat}) leads to
\begin{equation}
\label{mu_eq_zero_integral}
-\frac{3}{2}\int^1_{-1} \frac{1+R^2(1-\chi^2)}{\hat{\mu}_\text{eq}(0)
 + \chi}\chi\,d\chi = b^2.
\end{equation}
Thus, $\hat{\mu}_\text{eq}(0)$ may be approximated as
\begin{equation}
\label{mu_eq_zero_solution}
\hat{\mu}_\text{eq}(0)= \frac{1}{b}\sqrt{1+\frac{2 R^2}{5}} + {\cal
  O}(1)\qquad (b\rightarrow 0).
\end{equation}
Repeated integration by parts in (\ref{summand}) leads to
\begin{equation}
f_n = {\cal O}(\varepsilon^{3-2n})\qquad (\varepsilon\rightarrow
0),\qquad n\ge 2
\end{equation}
and we obtain the asymptotic approximation
\begin{align}
\notag
\int\limits^1_{-1}
\int\limits^{\hat{\mu}_{\text{eq}}}_{1+\varepsilon}
\bigg([\hat{\mu} + \chi]^2
&- \varepsilon^2[1 + R^2
  (1-\chi^2)]\bigg)^{1/2}\chi\,d\chi\,d\hat{\mu}\\
\notag
&= \frac{2}{3}(\hat{\mu}_\text{eq} - \hat{\mu}_\text{wb}) -
\frac{1}{2}\varepsilon^2\int\limits^1_{-1}
     [1+R^2(1-\chi^2)]\ln\bigg(\frac{\hat{\mu}_\text{eq}+\chi}{\hat{\mu}_\text{wb}+\chi}\bigg)\chi\,d\chi
+{\cal O}(\varepsilon^3)\qquad (\varepsilon\rightarrow 0)
\\
\label{approx_integral}
&= \frac{2}{3}(\hat{\mu}_\text{eq} - \hat{\mu}_\text{wb}) -
\frac{1}{2}\varepsilon^2\int\limits^1_{-1}
     [1+R^2(1-\chi^2)]\ln\bigg(\frac{\hat{\mu}_\text{eq}(0)+\chi}{1+\chi}\bigg)\chi\,d\chi
+{\cal O}(\varepsilon^3\ln\varepsilon)\qquad (\varepsilon\rightarrow 0).
\end{align}
Thus
\begin{equation}
\label{Emax_gourd_approx_mueq}
\frac{E^2_\text{max}}{a} =
mn_\text{ion}\varepsilon^2\bigg\{b^2(1-\hat{\mu}_\text{eq}(0)) +
  \frac{3}{2}\int\limits^1_{-1}
       [1+R^2(1-\chi^2)]\ln\bigg(\frac{1+\chi}{\hat{\mu}_\text{eq}(0)+\chi}\bigg)\chi\,d\chi\bigg\} + {\cal O}(\varepsilon^3\ln\varepsilon)\qquad (\varepsilon\rightarrow 0)
\end{equation}
and retaining lowest order terms in $\varepsilon$, $b$, $R$ yields
\begin{align}
\notag
E^2_\text{max} &\approx \frac{3}{2} mn_\text{ion}a\varepsilon^2\int\limits^1_{-1}
       [1+R^2(1-\chi^2)]\ln(1+\chi)\chi\,d\chi,\\
\notag
&= 2\pi\alpha m R^2\bigg(1+\frac{1}{3}R^2\bigg)\\
\label{Emax_gourd_approx_alpha}
&\approx \frac{3}{2}m n_\text{ion} \frac{1}{b}
\end{align}
where (\ref{def_b}) has been used to eliminate $\alpha R^2$.

\begin{figure}
\begin{center}
\scalebox{1.0}{\includegraphics{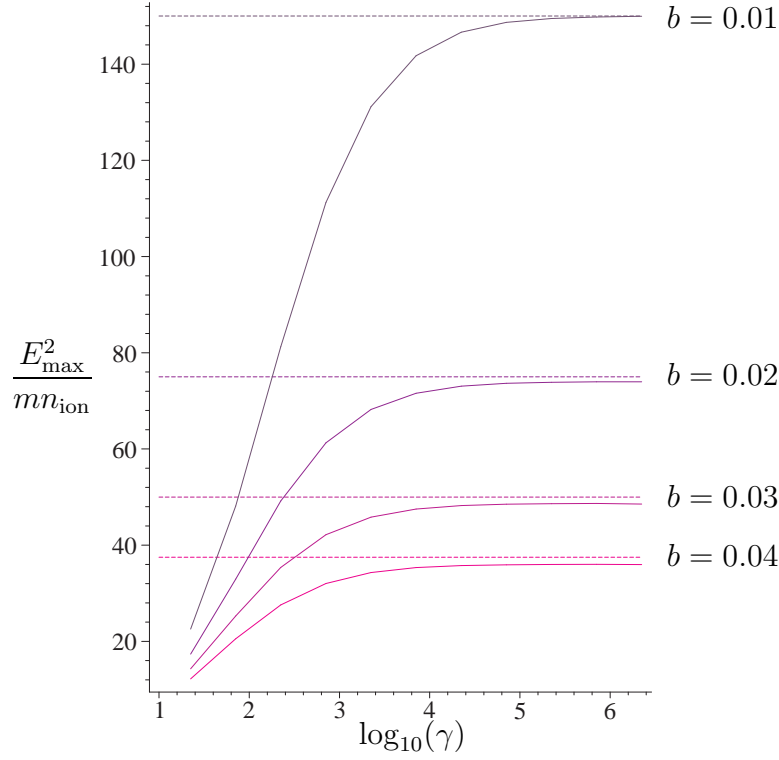}}
\caption{\label{figure:Emax2_vs_log10gamma}$E_\text{max}^2/(m n_\text{ion})$ versus
  $\log_{10}(\gamma)$ for $R=0.2$ and $b\in\{0.01,\,0.02,\,0.03,\,0.04\}$. The
dashed lines are the approximation (\ref{Emax_gourd_approx_alpha}) and the
solid curves are obtained by numerically integrating
(\ref{E_max_gourd_hat}-\ref{mu_wave-breaking_gourd_hat}).}
\end{center}
\end{figure}

Numerical validity of the above approximation is supported by
figure~\ref{figure:Emax2_vs_log10gamma}. The solid curves are obtained
by numerically integrating
(\ref{E_max_gourd_hat}-\ref{mu_wave-breaking_gourd_hat}) and the
dashed lines are obtained using (\ref{Emax_gourd_approx_alpha}).
It is clear that (\ref{Emax_gourd_approx_alpha}) yields a good approximation
to $E_\text{max}$ for large $\gamma$.

In order to compare (\ref{Emax_gourd_approx_alpha}) to expressions for
$E_\text{max}$ obtained elsewhere~\cite{katsouleas:1988, rosenzweig:1988, schroeder:2005}, it is useful to express
(\ref{Emax_gourd_approx_alpha}) as a function of effective
temperature. The electron proper number density $n=n_\text{ion}$ when
$\mu=\mu_{\text{eq}}$ and we eliminate $b$ in
favour of an effective longitudinal temperature $T_{\parallel\text{eq}}$ defined as 
\begin{equation}
\label{longitudinal_temp}
T_{\parallel\text{eq}} =
  \frac{1}{k_B\,n_{\text{ion}}}p_{\parallel\text{eq}}
\end{equation}
where $k_B$ is Boltzmann's constant and $p_{\parallel\text{eq}}$ is the
longitudinal pressure associated with the oscillator equilibrium $\mu=\mu_{\text{eq}}$. The
longitudinal pressure is
$p_{\parallel\text{eq}}=\mathcal{T}^{33}_\text{eq}$ where the
stress-energy-momentum tensor $\mathcal{T}_\text{eq}$ has components
\begin{equation}
\label{definition_stress_components}
\mathcal{T}^{ab}_\text{eq} = m\alpha \int_{\mathcal{W}_{\text{eq}}}
  \dot{x}^a\dot{x}^b
  \iota_X \# 1
\end{equation}
with $\mathcal{W}_{\text{eq}}$ the support of the waterbag distribution
$\mu=\mu_{\text{eq}}$ (the
choice of fibre is unimportant as the distribution associated with
$\mu_\text{eq}$ is independent of $\zeta$).

Since
\begin{equation}
\frac{\dot{x}^{32}\,d\dot{x}^1\wedge d\dot{x}^2 \wedge
d\dot{x}^3}{\sqrt{\beta^2 + \dot{x}^{32}}} =
d\bigg\{\bigg[\frac{1}{2}\dot{x}^3\sqrt{\beta^2+\dot{x}^{32}}
      -
      \frac{1}{2}\beta^2\sinh^{-1}\bigg(\frac{\dot{x}^3}{\beta}\bigg)\bigg]d\dot{x}^1\wedge d\dot{x}^2\bigg\}
\end{equation}
where $\dot{x}^{12} \equiv (\dot{x}^1)^2$, $\dot{x}^{22} \equiv (\dot{x}^2)^2$,
$\dot{x}^{32} \equiv (\dot{x}^3)^2$ and
$\beta\equiv\sqrt{1+\dot{x}^{12} + \dot{x}^{22}}$,
using (\ref{definition_stress_components}) and Stokes' theorem on forms, it follows
\begin{equation}
\label{longitudinal_stress}
\mathcal{T}^{33}_\text{eq} = m\alpha\int_{\partial
\mathcal{W}_{\text{eq}}} \bigg[\frac{1}{2}\dot{x}^3\sqrt{\beta^2+\dot{x}^{32}}
-\frac{1}{2}\beta^2\sinh^{-1}\bigg(\frac{\dot{x}^3}{\beta}\bigg)\bigg]d\dot{x}^1\wedge d\dot{x}^2.
\end{equation} 
Using (\ref{V_ansatz}, \ref{coframe}), components of the oscillator
equilibrium waterbag $\dot{x}^a = V^a_{\xi\,\text{eq}}$ are
\begin{align}
&\dot{x}^0 = \mu_\text{eq}-a\cos(\xi^1) - v\sqrt{[\mu_\text{eq} - a\cos(\xi^1)]^2 -
 \gamma^2[1 + R^2 \sin^2(\xi^1)]},\\
&\dot{x}^1 = R\sin(\xi^1)\cos(\xi^2),\\
&\dot{x}^2 = R\sin(\xi^1)\sin(\xi^2),\\
&\dot{x}^3 = v[\mu_\text{eq}-a\cos(\xi^1)] - \sqrt{[\mu_\text{eq} - a\cos(\xi^1)]^2 -
 \gamma^2[1 + R^2 \sin^2(\xi^1)]}
\end{align}
and it follows
\begin{align}
\notag
\frac{\dot{x}^3}{a} &= v[\hat{\mu}_\text{eq}-\cos(\xi^1)] - \sqrt{[\hat{\mu}_\text{eq} - \cos(\xi^1)]^2 -
 \varepsilon^2[1 + R^2 \sin^2(\xi^1)]}\\
\label{dotx3/a}
&\approx \bigg(-\frac{1}{2}b^2[\hat{\mu}_\text{eq}(0) - \cos(\xi^1)] +
\frac{1}{2[\hat{\mu}_\text{eq}(0) - \cos(\xi^1)]}\bigg)\varepsilon^2
\end{align}
to lowest order in $\varepsilon$ and $R$. Using
(\ref{mu_eq_zero_solution}, \ref{dotx3/a}) it follows
\begin{equation}
\frac{\dot{x}^3}{a} \approx \varepsilon^2 b^2\cos(\xi^1)
\end{equation}
to lowest order in $\varepsilon$, $b$ and $R$. Furthermore,
(\ref{def_varepsilon}, \ref{def_b}) yield $a\varepsilon^2 b^2 = b/v
\approx b$ and so
\begin{equation}
\dot{x}^3 \approx b\cos(\xi^1)
\end{equation}
to lowest order in $\varepsilon$, $b$ and $R$. Hence
\begin{equation}
\frac{\dot{x}^3}{2}\sqrt{\beta^2+\dot{x}^{32}}
-\frac{1}{2}\beta^2\sinh^{-1}\bigg(\frac{\dot{x}^3}{\beta}\bigg) \approx \frac{1}{3}b^3\cos^3(\xi^1)
\end{equation}
to lowest order in $\varepsilon$, $b$ and $R$ and (\ref{longitudinal_stress}) yields
\begin{align}
\notag
p_{\parallel \text{eq}} &\approx \frac{4\pi m\alpha R^2 b^3}{15}\\
\label{longitudinal_pressure}
&= \frac{1}{5}m n_\text{ion} b^2.
\end{align}
Equations (\ref{Emax_gourd_approx_alpha}, \ref{longitudinal_temp}, \ref{longitudinal_pressure}) yield 
\begin{equation}
\label{Emax2_gourd_final}
E_\text{max}^2 \approx \frac{m^2 \omega^2_p c^2}{q^2}\bigg(\frac{9 m c^2}{20
  k_B T_{\parallel\text{eq}}}\bigg)^{1/2},\qquad \varepsilon,b,R \ll 1
\end{equation}
where $\omega_p=\sqrt{n_\text{ion}q^2/(m\varepsilon_0)}$ is the plasma
frequency and the speed of light $c$ and permittivity of the
vacuum $\varepsilon_0$ have been restored.
\section*{Conclusion}
Equations (\ref{2d_Emax2_final}, \ref{Emax2_gourd_final}) indicate
that waterbags over $2$-dimensional spacetime and $4$-dimensional
spacetime can behave quite differently. Equation
(\ref{Emax2_gourd_final}) is independent of $\gamma$ but
(\ref{2d_Emax2_final}) diverges as $\gamma \rightarrow \infty$, and this
difference in behaviour arises because the logarithmic singularity in
the integrand in (\ref{Emax_gourd_approx_alpha}) is integrable.
Moreover, the $T_{\parallel \text{eq}}^{-1/4}$ behaviour of the asymptotic form
of $E_\text{max}$ for $k_B T_{\parallel \text{eq}}\ll mc^2$ is very
similar to the results of SES~\cite{schroeder:2005} and
others~\cite{rosenzweig:1988} in the limit
$v\rightarrow c$.

Direct comparison of our results and those of SES follows by setting the
transverse vector potential $\bm{A}_\perp$ to zero in the SES
model, thereby neglecting the overlap of the electromagnetic field of
the driver (laser pulse or particle bunch) and the wave. The approach followed by SES
begins with covariant field equations, induced from the Vlasov
equation, that couple the zeroth, first and second
order centred moments (in $\dot{x}^a$) of the $1$-particle
distribution $f$ with the electromagnetic
field. SES then assume that the $1$-particle distribution $f$
(restricted by pull-back to the unit hyperboloid) may be approximated as\footnote{We have changed the notation used by
SES to avoid conflict with our own.}
$f \simeq h(x^0,x^3,\dot{x}^3)\delta(\dot{x}^1)\delta(\dot{x}^2)$ where
$\delta$ is the Dirac delta function.
A covariant measure of the total thermal spread is given by the magnitude
$\epsilon^2$ of the ratio of the trace of the second order
centred moment and the zeroth moment. SES assume that the third
order centred moment is ${\cal O}(\epsilon^3)$ and can be neglected
relative to lower order moments. 

One could develop a similar argument to that given by SES based on
moments of a prescribed $3$-dimensional waterbag with narrow velocity
spread, rather than the line distribution employed by SES. However, nuances
in the shape of the waterbag would be lost; for example, we would not
know that merely the tip of the waterbag grazes the wave's phase velocity (see
figure~\ref{fig:gourd}) during the maximum amplitude oscillation. This
could be important because, as noted earlier, longitudinal
wave-breaking is associated with the trapping of considerable numbers
of particles in the wave (see~\cite{trines:2007} for a discussion),
and our present model neglects trapped particles. Thus, we expect that
$E_\text{max}$ calculated here is a lower bound on the maximum
electric field obtained when trapping is accounted for.

In conclusion, we have shown that it is possible to construct
$3$-dimensional waterbag distributions that lead to a maximum electric
field amplitude whose asymptotic behaviour is
similar to that of the SES model as $v\rightarrow c$ (with effective
temperature held fixed in the limit $v\rightarrow c$).
\section*{Acknowledgements}
We thank RMGM Trines for useful discussions. We acknowledge EPSRC for
financial support.


\begin{thebibliography}{99}
\small
\bibitem{tajima:1979}
T Tajima and JM Dawson, Phys. Rev. Lett. 43 (1979) 267
\bibitem{chen:1985}
P Chen, {\it et al.}, Phys. Rev. Lett. 54
(1985) 693
\bibitem{malka:2008}
V Malka, {\it et al.}, Nat. Phys. 4 (2008) 447 
\bibitem{caldwell:2009}
A Caldwell, {\it et al.}, Nat. Phys. 5 (2009) 363 
\bibitem{wuensch:2002}
W Wuensch, Proc. EPAC 2002 134 
\bibitem{schlenvoigt:2008}
HP Schlenvoigt, {\it et al.}, Nat. Phys. 4 (2008) 133 
%
\bibitem{akhiezer:1956} 
AI Akhiezer and RV Polovin, Sov. Phys. JETP 3 (1956) 696
\bibitem{dawson:1959} 
JM Dawson, Phys. Rev. 113 (1959) 383
\bibitem{coffey:1971} 
TP Coffey, Phys. Fluids 14 (1971) 1402
\bibitem{katsouleas:1988} 
T Katsouleas and WB Mori, Phys. Rev. Lett. 61 (1988) 90
\bibitem{rosenzweig:1988} 
JB Rosenzweig, Phys. Rev. A 38 (1988) 3634
\bibitem{schroeder:2005} 
CB Schroeder, E Esarey and BA Shadwick, Phys. Rev. E 72 (2005) 055401
\bibitem{trines:2006} 
RMGM Trines and PA Norreys, Phys. Plasmas 13 (2006) 123102
\bibitem{schroeder:2007} 
CB Schroeder, E Esarey and BA Shadwick, Phys. Plasmas 14 (2007) 084701
\bibitem{trines:2007} 
RMGM Trines and PA Norreys, Phys. Plasmas 14 (2007) 084702
\bibitem{mangles:2004} 
SPD Mangles, et al, Nature 431 (2004) 535--8
\bibitem{geddes:2004} 
CGR Geddes, et al, Nature 431 (2004) 538--41
\bibitem{faure:2004} 
J Faure, et al, Nature 431 (2004) 541--4
\bibitem{lu:2006} 
W Lu, {\it et al}, Phys. Plas. 13 (2006), 056709
\bibitem{esirkepov:2006} 
T Esirkepov, {\it et al.}, Phys. Rev. Lett. 96 (2006) 014803
\bibitem{mori:1990} 
WB Mori and T Katsouleas, Phys. Scr. T30 (1990) 127
\bibitem{burton_noble:2009}
DA Burton and A Noble, AIP Conf. Proc. 1086 (2009) 252
\bibitem{burton_noble_wen:2009}
DA Burton, A Noble and H Wen, Il Nuovo Cim. C 32 1 (2009) 1
\bibitem{ehlers:1971}
J Ehlers in {\it General Relativity and Cosmology}, Proceedings of the
International School of Physics ``Enrico Fermi'' 47, (Academic Press,
New York and London, 1971) 1
%
\bibitem{burton:2003}
DA Burton, Theoret. Appl. Mech. 30 (2003) 85
\bibitem{benn:1987}
IM Benn and RW Tucker, {\it An Introduction to Spinors and Geometry
  with Applications in Physics} (Adam Hilger, Bristol and New York, 1987)
\end{thebibliography}
\end{document}